\documentclass[a4,10pt]{article}
\usepackage{graphicx} 
\usepackage{booktabs} 
\usepackage{tabularx} 
\usepackage{caption}  
\usepackage{natbib} 
\usepackage{doi} 
\usepackage{authblk} 
\usepackage{rotating} 

\title{Percolation and Threshold-like Behavior in Multiple Sclerosis Progression}

\author[1]{Nikola Mirkov\thanks{Correspondence: nmirkov@vin.bg.ac.rs}}
\author[1]{Du\v{s}an S. Radivojevi\'{c}}
\author[1]{Slobodan Maleti\'{c}}

\affil[1]{'Vin\v{c}a' Institute of Nuclear Sciences - National Institute of the Republic of Serbia, University of Belgrade, Mike Petrovi\'{c}a Alasa 12-14, 11351 Belgrade, Serbia}

\date{}

\begin{document}

\maketitle

\begin{abstract}
    In this study we investigate the Percolation Hypothesis for Multiple Sclerosis Progression. The methodology relies on cross-reference analysis centered around a question: \textit{What is the evidence for a Percolation/phase-transition hypothesis in Multiple Sclerosis (MS), especially the idea that the RRMS dynamic balance can abruptly break akin to crossing a percolation threshold into SPMS?} We identify theoretical models invoking percolation/critical thresholds, network/connectome studies assessing percolation robustness or threshold-like behavior, clinical markers showing thresholds or early-warning signals, and counter-evidence arguing for gradual/continuum transitions. 
\end{abstract}

\section{Introduction}

It has recently been pointed out (\cite{Papo2026}) that in addition to brain function, brain dysfunction can also be related to its representation as a complex network. Such a perspective enables notions and concepts form complex network theory, and statistical mechanics to be potentially useful in diseases where the brain and the CNS can be modelled as the complex network of neurons \cite{sporns2011, chialvo2010, bassett1, bassett2}.\\
The intersection of complex networks and statistical mechanics provides a theoretical framework beyond static anatomical maps to neural systems as dynamic entities where macroscopic properties (like cognition or consciousness) emerge from microscopic interactions. Statistical mechanics supplies the tools to analyze how global system behavior arises from the collective dynamics (of billions of neurons) capable of undergoing phase transitions (\cite{cohen2016}).\\
A central concept of percolation theory\footnote{In physics and mathematics, percolation is classified as a geometric phase transition driven by connectivity and spatial occupancy.} models the formation of connected clusters within a network (\cite{saberi2015}). 
Percolation can describe how neural signals propagate across the structural connectome; if connectivity drops below a specific percolation threshold\footnote{Percolation Threshold ($p_c$)-The critical occupancy probability at which a system transitions from many small, isolated clusters to one "giant" connected component.} the network fragments, halting global information transfer. Conversely, excessive connectivity can lead to pathological hypersynchrony (\cite{palop2017}). Evidence suggests that a healthy brain operates near a critical point at the edge of this transition-a state of "criticality"\footnote{The state of a system precisely at a phase transition, characterized by extreme sensitivity to small structural or parameter changes.}-which maximizes dynamic range, sensitivity to stimuli, and the ability to rapidly reconfigure network topology in response to complex tasks \cite{chialvo2004, miesel2012, scheffer2009}.\\
Multiple Sclerosis (MS) can be effectively modeled within this framework as a progressive process of network fragmentation (\cite{schoonheim2022}). In the language of statistical mechanics and complex networks, MS pathology (demyelination and axonal loss) functions as a dynamic process of edge removal or node failure within the structural connectome. Redundancy which maintains global communication despite localized damage exhibits significant resilience. As damage accumulates, the network efficiency degrades non-linearly.
Percolation theory can particularly be relevant for explaining the clinical course of the disease: the system resists functional collapse up to a certain critical density of damage. In that picture MS progression can be viewed as the network approaching a percolation threshold. Once this threshold is crossed, the giant component (the globally connected functional cluster) disintegrates, leading to a rapid onset of severe cognitive or physical disability. \\
Complex networks and percolation help address the Clinico-Radiological Paradox by explaining why a patient's physical lesion load (MRI visible damage) often does not correlate perfectly with their disability, using the idea of topological vulnerability. The location of the damage within the network topology (e.g., hitting a highly connected "hub" vs. a peripheral node) matters more than the sheer volume of damage. The brain's backbone of highly connected hubs (the "rich-club" network\footnote{A network topology where high-degree "hubs" are more densely interconnected with each other than with low-degree peripheral nodes}) such as the precuneus, superior frontal cortex, and insula is specifically affected in MS, and its degradation is a key driver of cognitive decline. In MS, this backbone is affected through a mechanism often described as Hub Overload followed by Network Collapse.\\
Because MS lesions (demyelination) appear somewhat randomly in the white matter, the brain initially compensates by rerouting information flow around the damage. Since the "rich-club" hubs are central to almost all global information transfer, this rerouting forces excessive traffic through the remaining healthy hubs (\cite{cao2020}). These hubs are effectively forced to operate at a hyper-metabolic rate to maintain network integration. This constant state of over-activation is strongly linked to cognitive fatigue, a hallmark symptom of MS. The brain is working harder to achieve the same connectivity (\cite{stam2024}). The rich-club relies heavily on long-range white matter tracts to connect distant brain regions (e.g., front-to-back connectivity). These long tracts have a higher statistical probability of being hit by a lesion simply because they traverse more spatial volume, causing vulnerability. As MS progresses, a measure of how well these hubs are connected to each other (rich-club coefficient), drops significantly. When these specific links are severed (Disconnection of Long-Range Highways), the brain loses its ability to integrate information globally, leading to slowed information processing speed (\cite{stellmann2017}). The Network Collapse threshold explains the non-linear progression of disability in MS. Two states are in action. The Compensated State: The rich-club holds the network together despite peripheral damage (high resilience). The Collapsed State: Once the damage density hits a critical threshold (percolation threshold), the rich-club itself fragments. The network shifts from a globally integrated system to a segregated one. This transition point often marks the onset of progressive cognitive impairment that is irreversible.\\
In this paper, we investigate whether the progression of Multiple Sclerosis (MS), especially the transition from relapse–remitting MS (RRMS)\footnote{An MS course defined by clearly identifiable attacks of new symptoms, followed by periods of partial or complete recovery.} to secondary progressive MS (SPMS)\footnote{A stage of MS following RRMS characterized by a steady accumulation of neurological disability independent of distinct relapses.}, can be framed as a percolation/phase‑transition phenomenon. Below we synthesize supporting and opposing evidence across theory, cognitive networks, connectomics, and clinical frameworks.\\
Our first objective is to clarify the percolation/phase-transition hypothesis in Multiple Sclerosis (MS). To build a conceptual framework we first need to address what is meant by percolation threshold in RRMS dynamics and how it could map to biological/clinical observables (lesions, connectivity, immune networks). \\
The hypothesis posits the following.
\emph{In RRMS, inflammatory/demyelinating events stochastically damage nodes/edges in CNS networks (white matter tracts, functional connectivity). A percolation threshold would be reached when lesion/disconnection density crosses a critical value causing a giant-component collapse and abrupt, persistent disability progression (RRMS to SPMS).} \\
The questions arising from this definition are: (1) Are there data indicating critical slowing/early-warning signals before SPMS? (2) Do structural/functional connectomes in MS show percolation-like robustness–fragility curves? (3) Can lesion topology and axonal reserve map to percolation parameters (p, pc)? (4) Do immune network activation patterns show percolation/avalanche statistics? (5) What clinical markers (EDSS\footnote{The Expanded Disability Status Scale; a standardized 0–10 point system used to quantify and track Multiple Sclerosis disease progression.}, brain atrophy, NfL) exhibit threshold behavior?

\subsection{Plan and Scope}

In the sequel, we search the literature (papers, reviews, theory) linking percolation theory or phase transitions to MS and RRMS to SPMS transitions. Primary evidence is needed to see whether percolation models have been proposed/applied to MS, and which data support them. Extracting and synthesizing key evidence (methods, datasets, findings) including mathematical/biophysical models and clinical correlates (lesion load, connectivity, biomarkers). Gathering detailed support or counter-evidence, including mechanistic mapping and predictions of the percolation hypothesis. Assessing alternative explanations and limitations (e.g., gradual neurodegeneration models) to contextualize the percolation hypothesis. Weighing competing models and identifying testable discriminators. Summarizing evidence for or against the percolation hypothesis.

\section{Evaluating Hypothesis through Cross-Study Synthesis}

Primary evidence is needed to see whether percolation models have been applied to MS, and which data support them. The survey included searching for joint mentions of percolation with multiple sclerosis along with RRMS, SPMS, and phase transition.\\
Diverse sources are found spanning theory, network percolation in cognition, connectomics, and clinical conceptual models. We have identified: (a) Theoretical threshold/phase-transition model of MS with explicit thresholds and phase diagram (\cite{kannan2017aminimalunified}). (b) Cognitive-network percolation analyses showing reduced percolation robustness in MS semantic networks (\cite{lebkuecher2024, lall2024}). (c) Connectome studies showing phenotype-dependent connectivity degradation and hub-centric disconnection, but no explicit percolation tests (\cite{pagani2020, martinezheras2023, tranfa2024}). (d) Conceptual threshold models of reserve/decompensation and transitional phase frameworks (\cite{vollmer2021, kleiter2020}).
Following key strands within these sources, linking thresholds/percolation across theory, cognitive networks, connectomics, and clinical concepts with balanced counterpoints are distinguished: \\
(1) Theory - \cite{kannan2017aminimalunified} introduces explicit immune and CNS thresholds; crossing them converts oscillatory RRMS-like dynamics to monotone progressive dynamics and yields a phase diagram of MS phenotypes. \\
(2) Cognitive networks - Percolation analyses on semantic networks show MS networks fragment faster (lower percolation integrals) and require higher clique-intensity thresholds for giant component emergence than controls, indicating reduced robustness (\cite{lebkuecher2024, lall2024}). \\
(3) Connectome-Structural connectivity studies demonstrate phenotype-dependent network degradation and hub-centric disconnection; SPMS shows widespread reductions compared with CIS/RRMS, but explicit percolation thresholds are not tested (\cite{martinezheras2023, tranfa2024}). \\
(4) Disconnection modeling, \cite{pagani2020} suggests resilience to observed lesion disconnections, arguing against simple abrupt collapse in typical lesion patterns. \\
(5) Clinical frameworks - Topographical threshold/recapitulation model (Laitman/Krieger) conceptualizes reserve and threshold emergence of deficits (\cite{laitman2018thetopographicalmodel}). Continuum/reserve models propose decompensation when reserve falls below a critical level (\cite{vollmer2021}). Transitional-phase reviews emphasize operational thresholds and candidate early markers, but highlight heterogeneity and measurement limits (\cite{kleiter2020}).\\
Regarding alternative explanations continuum/reserve models posit gradual loss of neurologic reserve with phenotypes along a spectrum; pathology overlaps between relapsing and progressive MS and biomarkers poorly separate them (\cite{vollmer2021}). Transitional-phase frameworks caution against sharp event-based thresholds; measurement noise and heterogeneity limit abrupt definitions (\cite{kleiter2020}). Connectome disconnection modeling indicates resilience under observed lesions (\cite{pagani2020}), arguing against simple global percolation collapse in typical cases. 

\section{Percolation Evidence Summary}

A coherent theoretical percolation/threshold framework exists and cognitive network percolation analyses show empirically reduced robustness in MS semantic networks. Structural connectome work demonstrates progressive, hub‑centred disruption but, using typical lesion distributions, finds substantial resilience and does not yet provide a direct empirical demonstration of a global percolation threshold governing RRMS to SPMS conversion. Clinical and review literature provide conceptually-compatible threshold analogies (reserve/topography) but emphasize heterogeneity and measurement limits that complicate claims of a universal abrupt phase transition.\\
Existing theoretical support: A minimal dynamical model (\cite{kannan2017aminimalunified}) explicitly defines immune and CNS thresholds whose crossing produces irreversible progressive dynamics, formalizing the phase‑transition idea for RRMS to SPMS. \\
Cognitive‑network percolation: Semantic networks in MS display reduced percolation robustness and faster fragmentation, indicating vulnerability to threshold-like breakdown in information processing networks; this supports the percolation analogy at a cognitive‑network level but does not prove a disease‑wide percolation threshold (\cite{lebkuecher2024, lall2024}). \\
Connectome and lesions: Large, multisite and longitudinal connectomics show phenotype‑dependent degradation and hub‑centric disconnection linked to disability, yet empirical modeling of lesion‑induced disconnection often indicates network resilience under typical lesion distributions, arguing against a simple, universal percolation collapse mechanism (\cite{ martinezheras2023, tranfa2024, pagani2020}). \\
Clinical course: Reserve/topography models align with a threshold/decompensation analogy and are consistent with abrupt‑appearing clinical transitions, but comprehensive reviews emphasize that MS likely unfolds along a continuum with heterogeneous trajectories and measurement noise that complicate defining a sharp, universal threshold (\cite{laitman2018thetopographicalmodel, vollmer2021, kleiter2020}).

Table \ref{tab:research-summary} summarizes percolation evidence across theory, cognitive networks, connectomics, and clinical frameworks with predictions. See Appendix A.I and A.II for detailed discussion on findings.

\begin{sidewaystable}[ph!]
    \centering
    \caption{Summary of Evidences, Key Findings and Testable predictions related to the Percolation Hypothesis.}
    \label{tab:research-summary}
    \begin{tabularx}{\textwidth}{XXXXX} 
        \toprule
        \textbf{Evidence type} & \textbf{Key finding} & \textbf{Supports
        percolation / phase‑transition?} & \textbf{Methods / dataset} & \textbf{Testable prediction or implication} \\ 
        \midrule
        Minimal dynamical model with explicit thresholds & A two-threshold (immune $I_C$ and CNS $Z_C$) model produces RRMS‑like oscillations while noise‑driven crossing of thresholds yields irreversible progressive dynamics and a phase diagram of MS phenotypes. Authors emphasize loss of negative feedback as a switch to progression. & Yes (theoretical, mechanistic) & Reduced, coarse‑grained dynamical model; stochastic simulations across parameter sweeps \cite{kannan2017aminimalunified} & If true, one should observe abrupt changes in system‑level markers (e.g., persistent rise in inflammatory proxy or irreversible increase in damage markers) tied to early‑warning signatures before the switch. \\
        \addlinespace 
        Cognitive (semantic) network percolation analyses & Semantic networks from pwMS fragment faster under clique-percolation pruning (lower percolation integrals) and require higher clique‑intensity to form a giant component; spreading‑activation decays faster in MS networks. & Partial-direct percolation measures show reduced robustness in cognitive networks, consistent with threshold vulnerability in information networks & Percolation / clique‑percolation on semantic fluency-derived networks; percolation integrals and spreading‑activation simulations \cite{lebkuecher2024, lall2024} & Network fragility at cognitive level predicts faster cognitive decline once a critical subset of semantic links/nodes are lost; test by longitudinal semantic network percolation and relating percolation slope to cognitive deterioration. \\
        \addlinespace 
        Structural connectome comparisons across phenotypes & SPMS shows more widespread reductions in FA and global/local graph metrics than CIS/RRMS; disconnection concentrated around thalamic and hub regions and predicts disability progression moderately. & Mixed - indicates progressive, non‑linear worsening of network metrics (step‑like across phenotypes) but does not directly demonstrate a single sharp percolation threshold & Diffusion MRI tractography, FA‑weighted connectomes, NBS and longitudinal conventional MRI‑based disconnection mapping  \cite{martinezheras2023, tranfa2024} & If percolation applies, lesion/disconnection topology (targeting hubs or rich‑club) will produce abrupt functional collapse; simulate targeted vs random lesioning on subject connectomes and compare to clinical transitions. \\
        
        \bottomrule
    \end{tabularx}
\end{sidewaystable}

\begin{sidewaystable}[ph!]
    \centering
    \caption*{Table 1 (Continued): Summary of Evidences, Key Findings and Testable predictions related to the Percolation Hypothesis.}
    \begin{tabularx}{\textwidth}{XXXXX} 
        \toprule
        \textbf{Evidence type} & \textbf{Key finding} & \textbf{Supports
        percolation / phase‑transition?} & \textbf{Methods / dataset} & \textbf{Testable prediction or implication} \\ 
        \midrule
        Modeled lesion disconnection resilience & Modeling lesion‑induced disconnection on empirical lesion sets produced only modest network changes-interpreted as brain resilience to typical lesion distributions, arguing against a simple global collapse. & No (empirical modeling argues against simple global percolation collapse under observed lesion patterns) & Tractography templates + lesion overlap → modeled disconnection maps; comparison of network metrics before/after modeled disconnection \cite{pagani2020} & Predicts that typical, spatially dispersed lesion loads produce gradual functional loss; percolation‑like abrupt collapse would require either different lesion topologies (e.g., hub‑targeting) or additional mechanisms (axonal degeneration, compartmentalized inflammation). \\
        \addlinespace
        Topographical / reserve clinical models & Disability emerges as lesions or prior relapse loci become unmasked when neurologic/functional reserve (the 'water level') falls below a threshold; the model can flag earlier loci of future progression. & Conceptually supports threshold/decompensation view (phenomenological) & Clinical topographical mapping and longitudinal symptom mapping; recapitulation hypothesis testing in small cohorts \cite{laitman2018thetopographicalmodel} & Longitudinal mapping should detect earlier re‑emergence of prior relapse deficits as reserve declines - an empirical early‑warning signature consistent with threshold crossing. \\
        \addlinespace
        Continuum / reserve and transitional‑phase reviews & MS phenotypes better conceptualized as a continuum driven by progressive depletion of reserve and aging; diagnostic thresholds are operational and noisy; biomarkers overlap across phases. & No / Alternative: favors gradual depletion rather than single abrupt phase change in many cases & Reviews synthesizing pathology, imaging and biomarker literature; emphasize heterogeneity and measurement limitations \cite{vollmer2021, kleiter2020} & Distinguishing gradual vs abrupt models requires sensitive longitudinal biomarkers (e.g., high‑frequency imaging, continuous sensor measures, NfL dynamics) to detect critical slowing or discrete jumps. \\
        \bottomrule
    \end{tabularx}
\end{sidewaystable}

\subsection{Testable predictions}

As noted, a plausible theoretical mechanistic percolation/phase‑transition mechanism maps to immune‑ and CNS‑threshold variables and predicts noise‑triggered switches from RRMS to SPMS. On the other hand the empirical network evidence is mixed - cognitive (semantic) networks show formal percolation fragility consistent with threshold vulnerability at the functional/cognitive level (\cite{lebkuecher2024, lall2024}), while structural connectome studies show progressive, phenotype‑dependent degradation but have not directly demonstrated a global percolation threshold in typical lesion distributions (\cite{martinezheras2023, tranfa2024, pagani2020}). Clinical frameworks support a threshold/decompensation analogy via loss of reserve and topographical unmasking (\cite{laitman2018thetopographicalmodel, vollmer2021}), but authors caution heterogeneity and operational/measurement limits that can masquerade as abrupt transitions (\cite{kleiter2020}).

The synthesized findings underscore the need for further research capable of elucidating the factors underlying these results. The following experiments can be used to further discriminate models:
\begin{itemize}
    \item Longitudinal percolation analysis on subject‑level structural/functional connectomes: compute percolation curves and percolation integrals over time and test for sudden slope changes preceding clinical SPMS conversion (prediction: true percolation model to early‑warning signals, critical slowing, sudden slope change);
    \item Targeted in‑silico lesioning: compare simulated random vs hub‑targeted lesioning on individual connectomes to determine whether observed lesion topologies could produce abrupt giant‑component collapse (prediction: abrupt collapse requires hub‑targeting or compounded processes like compartmentalized axonopathy);
    \item High‑frequency biomarker monitoring (serial NfL, sensitive imaging, continuous gait/cognitive sensors) to look for critical slowing, rising variance, or shifts in recovery dynamics before irreversible progression (prediction: percolation/bifurcation model to early‑warning statistical signatures). 
\end{itemize}

\subsection{Going back to Initial Questions}

After gathering the relevant evidence, we attempt to explicitly answer the questions posed in the introduction.

\emph{Q: Are there data indicating critical slowing/early-warning signals before SPMS?}\\
\emph{Answer: Partial.} Reviews of the transitional phase emphasize that pathology, imaging, and biomarkers evolve continuously and that no standardized imaging measure prospectively predicts conversion on an individual basis; nevertheless, group‑level MRI and biofluid markers (gray‑matter and deep gray‑matter atrophy, cortical lesion burden) and serum neurofilament light (sNfL) elevations anticipate future disability and brain/spinal atrophy, consistent with pre‑ progression deterioration, though not formal critical‑slowing statistics. Proposed early indicators include objective MSFC changes and continuous sensor metrics during relapse‑free intervals, but explicit calculations of autocorrelation, variance inflation, or recovery‑time “critical slowing” are not reported in the cited texts (\cite{kleiter2020, filippi2020}).

\emph{Q: Do structural/functional connectomes in MS show percolation‑like robustness–fragility curves?}\\
\emph{Answer: Partial.} Longitudinal and cross‑sectional connectomics show distributed subnetworks of structural disconnection and morphometric similarity disruption associated with EDSS and SDMT, with modest ability of baseline disconnection to predict long‑term confirmed disability progression. These works apply thresholded graph analyses and network‑based statistics but do not report explicit percolation curves relating edge‑occupancy to giant‑component collapse. Thus, connectome degradation and hub‑centric vulnerability are documented without demonstration of formal percolation robustness curves (\cite{tranfa2024}).

\emph{Q: Can lesion topology and axonal reserve map to percolation parameters (p, pc)?}\\
\emph{Answer: Partial.} Structural disconnection correlates strongly with total lesion volume and identifies progressive subnetworks (e.g., fronto‑thalamic), and relationships between disconnection and morphometric similarity can be nonlinear/multiphasic-features compatible with threshold phenomena. However, no study explicitly estimates an edge‑occupancy p or critical percolation threshold pc from lesion topology or reserve metrics in vivo; current results provide qualitative support but not parameterized mappings (\cite{tranfa2024}).

\emph{Q: Do immune network activation patterns show percolation/avalanche statistics?}\\
\emph{Answer: No (within available MS evidence).} The summarized transitional‑phase and biomarker reviews document compartmentalized/chronic inflammation but do not report avalanche or power‑law statistics of immune activation specific to MS; such analyses are a present gap in the provided sources (\cite{kleiter2020, filippi2020}).

\emph{Q: What clinical markers (EDSS, brain atrophy, NfL) exhibit threshold behavior?} \\
\emph{Answer: Partial evidence.} sNfL is elevated in progressive phenotypes and predicts near‑term EDSS worsening and longer‑term brain and spinal cord atrophy; GM and deep GM atrophy rates rise with progression; higher cortical lesion burden at onset ($> 7$ lesions in one study) was associated with faster conversion to SPMS. However, reviews stress that individual‑level threshold cut‑offs and standardized pathological cut‑points remain undefined, and EDSS rises confirmed over short windows can overestimate permanent disability; MSFC‑based deterioration may better flag relapse‑independent progression operationally but requires further validation (\cite{kleiter2020, filippi2020}).

Table \ref{tab:research-summary-2} presents a summary of these results. Noting the key limitations helps map current empirical gaps to percolation-theoretic concepts.

\begin{sidewaystable}[ph!]
    \centering
    \caption{A summary table of answers to five percolation‑related questions: each row states whether evidence supports (Yes/No/Partial), gives a concise evidence summary, notes main limitations, and cites the sources used. }
    \label{tab:research-summary-2}
    \begin{tabularx}{\textwidth}{XXXXX} 
        \toprule
        \textbf{Question} & \textbf{Answer} & \textbf{Evidence sumamry} & \textbf{Key limitations} & \textbf{Sources} \\ 
        \midrule

   1) Critical slowing / early-warning before SPMS? & Partial & No studies report formal critical‑slowing metrics (autocorrelation, recovery time) pre‑SPMS; longitudinal biomarkers (rising sNfL, accelerating GM/deep GM atrophy) show signals years before conversion but are associative rather than classical early‑warning statistics. & Lack of high‑temporal‑resolution analyses and explicit early‑warning indicator computations. & \cite{filippi2020, kleiter2020}  \\
   \addlinespace
   
   2) Connectome percolation robustness–fragility curves? & Partial & Structural/functional network studies identify subnetworks, rich‑club/hub weakening and thresholded analyses (density thresholds) with modest prediction of progression; however, they do not present formal percolation/giant‑component vs. edge‑occupancy curves. & No explicit percolation‑curve analyses (p to giant component) reported. & \cite{tranfa2024, filippi2020} \\
   \addlinespace
   
        \bottomrule
    \end{tabularx}
\end{sidewaystable}

\begin{sidewaystable}[ph!]
    \centering
    \caption*{Table 2 (Continued).}
    \begin{tabularx}{\textwidth}{XXXXX} 
        \toprule
        \textbf{Question} & \textbf{Answer} & \textbf{Evidence sumamry} & \textbf{Key limitations} & \textbf{Sources} \\ 
        \midrule

   3) Map lesion topology / axonal reserve to percolation parameters (p, pc)? & Partial & Structural disconnection correlates strongly with total lesion volume and identifies progressive subnetworks; authors note nonlinear/multiphasic edge‑level relationships suggestive of thresholded network collapse potential. & No direct mapping of lesion maps or axonal reserve to quantitative p or estimated pc values. & \cite{tranfa2024} \\
   \addlinespace
   
   4) Immune network activation showing percolation / avalanche statistics? & No (within available MS data) & Pathology and imaging document chronic active lesions and compartmentalized inflammation in the transitional phase, but MS‑specific immune avalanche / power‑law avalanche analyses were not identified in the reviewed texts. & Absence of MS datasets analysed for avalanche/percolation statistics; theoretical/experimental gap. & \cite{kleiter2020, filippi2020} \\
   \addlinespace
   
   5) Clinical markers (EDSS, brain atrophy, NfL) exhibiting threshold behavior? & Partial & sNfL is elevated in progressive phenotypes and predicts EDSS worsening and future brain/spinal atrophy; GM and deep GM atrophy rates rise markedly with progression; EDSS/MSFC changes detect group‑level transitions but have sensitivity/standardization limits. & Biomarker signals are robust at group level but individual threshold cutoffs and prospective validation for single‑patient prediction remain lacking. & \cite{kleiter2020, filippi2020, tranfa2024} \\

        \bottomrule
    \end{tabularx}
\end{sidewaystable}

Across all five questions, current evidence supports deterioration consistent with threshold‑like processes (especially in biomarkers and subnetworks), but explicit demonstrations of dynamical critical slowing, percolation robustness curves, or quantitative mappings to percolation parameters in vivo are lacking in the cited texts (see Appendix A.III for details). Larger longitudinal datasets with high‑frequency measurements and explicit early‑warning and percolation analyses are needed to adjudicate the percolation hypothesis more directly (\cite{kleiter2020, filippi2020, tranfa2024}).

\section{Conclusion}

A percolation/phase‑transition hypothesis for MS is theoretically plausible and supported by cognitive‑network percolation findings, but direct, disease‑wide empirical evidence of a global percolation threshold governing RRMS to SPMS is lacking. Current connectomic and clinical data favour a continuum with potential domain‑specific thresholds (e.g., hub‑targeted disconnection or reserve exhaustion) rather than a single universal critical point. Prospective longitudinal percolation analyses of individual connectomes, targeted lesioning simulations, and high‑frequency biomarker monitoring are needed to decisively test for early‑warning signatures and abrupt transitions. The state of the hypothesis is inconclusive, and more research in this direction is needed.

\bibliographystyle{plainnat}
\bibliography{references} 

\newpage
\section*{Appendix A}

\subsection*{I Can RRMS dynamic balance abruptly break into SPMS?}

We ask what is the evidence for a Percolation/phase-transition hypothesis in Multiple Sclerosis (MS), especially the idea that the RRMS dynamic balance can abruptly break akin to crossing a percolation threshold into SPMS? 


Kannan et al. \cite{kannan2017aminimalunified} present a minimal unified mathematical model of MS that constructs a two-dimensional phase diagram of disease trajectories by sweeping parameter space and running many stochastic realizations. They explicitly classify trajectories as RRMS when no threshold is crossed and as SPMS/PPMS when either immune or CNS thresholds are breached, and determine phase boundaries by thresholding the fraction of realizations (e.g., using 95\%/5\% criteria). The model emphasizes interaction structure (cross-regulation between inflammatory and anti-inflammatory components) and oscillatory immune dynamics rather than specific biochemical details. The authors argue that relapse–remit behavior arises from periodic/quasi-periodic immune dynamics and that transition to progressive disease occurs when the suppressive negative-feedback fails (an abrupt threshold-like breakdown). They abstract immune actors (T/B cells, microglia, cytokines) into generic inflammatory vs suppressive components, noting that resistance of inflammatory cells to regulation has been observed empirically. Overall the paper provides direct theoretical support for a threshold/phase-transition interpretation of RRMS to SPMS transitions via numerical phase diagrams and threshold definitions.

\cite{kannan2017aminimalunified}  also present a minimal dynamical model of MS in which two explicit thresholds govern the qualitative behavior of the disease: an immune threshold $I_C$ and a CNS threshold $Z_C$. When the inflammatory variable $I$ remains below $I_C$ the system orbits a stable fixed point and shows oscillatory (relapse–remit) dynamics; stochastic perturbations can push $I$ past $I_C$, at which point the model “severs” a negative feedback loop and the inflammatory component instead increases monotonically (a switch to progressive disease/SPMS). The model also includes a CNS demyelination variable $Z_{Demy}$ with remyelination (rate k) and a CNS threshold $Z_C$: when $Z_{Demy} > Z_C$ neurodegeneration ($Z_{Dead}$) is triggered and damage accumulates. The authors keep all parameters constant except the two thresholds to demonstrate that varying thresholds alone can produce RRMS, SPMS, PPMS and PRMS phenotypes. Noise-driven crossings of thresholds are emphasized as critical in determining if/when abrupt transitions occur. Thus the paper provides a theoretical threshold/bifurcation mechanism (noise-induced threshold crossing leading to loss of oscillatory attractor) consistent with a phase-transition-like switch from RRMS to SPMS.

\cite{kannan2017aminimalunified} thus propose a minimal, coarse-grained computational model of MS that frames the RRMS to SPMS transition as a threshold-driven change in system dynamics. The authors explicitly state the hypothesis that "the transition from RRMS to SPMS occurs when the extent or nature of injury reaches a certain threshold" and construct a model "involving the immune system and CNS" that "generates the principal sub-types of the disease." The model introduces two explicit capacities/thresholds (one immune, one CNS) that "separate dynamically distinct behavior of the model." One critical event is described as "the collapse of the negative feedback loop in the immune system when the inflammatory component reaches a certain threshold," after which inflammation escalates irreversibly. The paper also emphasizes that stochastic fluctuations ("noise") are important in determining whether/when thresholds are crossed, and that scanning the two-dimensional threshold space yields "multiple phases of disease evolution" and substantial heterogeneity beyond standard clinical classes. The model supports a phase-transition/percolation-like framing for MS progression but is a theoretical, minimal model and does not provide empirical connectome/percolation analyses or specific clinical early-warning markers.

\cite{laitman2018thetopographicalmodel} introduce the 'topographical model' of MS which explicitly frames clinical progression in threshold-like terms. The CNS is visualized as a pool whose water depth encodes regional functional reserve; lesions are topographical peaks that may rise above a clinical threshold to cause symptoms. As 'functional reserve' (the water level) declines, previously subthreshold lesions can become clinically manifest, so that the "combined volume of above-threshold topographical peaks corresponds with the degree of accumulated disability." The authors label this the 'recapitulation hypothesis': prior relapse signs and lesion localization could predict individualized patterns of later disability as reserve is lost. They implement the model in disease-simulation software to search for insidious reemergence of prior relapse symptoms as potential early-warning or heralding signs of transition from RRMS to SPMS. The paper empirically tests this idea in a small longitudinal subgroup (14 patients identified with diagnostic uncertainty; 10 had sufficient records). The model is explicitly conceptual and "agnostic to specific pathophysiological and cellular mechanisms," offering a clinical/phenomenological threshold framework rather than a mechanistic percolation model.

 The study by \cite{giovannoni2017} advances a length-dependent central-axonopathy framing of MS in which axonal length, anatomical architecture and stochastic "statistical phenomena" interact to determine when pathways fail. Shorter axons (sensory relays) may be relatively protected, while long projection motor axons are more likely to accumulate multiple 'hits' and exhaust functional reserve, producing worse outcomes. This underpins the "sensory-motor paradox" (sensory relapses more frequent early; motor attacks worse and more common later). The authors propose an "asynchronous progressive MS hypothesis": different neuronal domains have different disease-course time windows, so some systems may already be exhausted and progressive while others retain reserve and remain modifiable. The text implies threshold-like behavior, once a neuronal system "has exhausted its reserve capacity" further decline may be hard to alter, and attributes stochasticity to which pathways become symptomatic early. However, the study contains no formal percolation or phase-transition models, no network/connectome robustness analyses, and no explicit clinical early-warning signal data; it suggests conceptual support for abrupt or domain-specific transitions but lacks quantitative or connectomics evidence.

 The review by \cite{vollmer2021} frames MS as a continuum from relapsing (RMS) to progressive (PrMS) disease and proposes neurologic (brain) reserve as the key buffering mechanism. Early inflammatory activity causes brain atrophy whose clinical effects are initially compensated by reserve; continued loss from MS and aging depletes reserve and then 'unmasks' subclinical disease and aging effects, producing progressive disability. The authors state neuronal/brain volume loss is increasingly viewed as the main driver of disability and note that progressive vs relapsing disease cannot currently be distinguished by diagnostic tests. They argue phenotypic differences arise from differing levels of reserve rather than distinct biological entities, and that loss of reserve explains why patients with PrMS cannot recover function and respond less to therapies. While this conceptual model implies a threshold-like transition (buffered state $\rightarrow$ decompensation when reserve falls below a critical level), the study contains no explicit percolation/phase-transition terminology, no mathematical or network/connectome models, no empirical threshold values or early-warning signal analyses, and no direct counter-evidence. Thus the paper provides conceptual support for a threshold/depletion mechanism but lacks mechanistic percolation modeling or empirical threshold markers.

They (\cite{vollmer2021}) frame MS phenotypes as a continuum and presents the neurologic/brain reserve concept as a mechanism that produces threshold-like change: individuals with larger brain reserve (e.g., larger ICV, cognitive reserve) can buffer subclinical inflammatory injury, delaying symptom expression, whereas progressive disease emerges as reserve is exhausted. Age and disease duration correlate with transition to progressive MS; brain atrophy begins early and accelerates with aging, and by $\sim60$ years $>50\%$ of observed brain loss may be due to normal aging rather than MS. The authors note that no immunologic or biomarker test clearly discriminates relapsing from progressive MS and that NfL is confounded by age. They cite studies concluding pathology is not different between relapsing and progressive MS, and point out small-study limitations and immunosenescence as possible confounders. Overall, the text supports a threshold/continuum interpretation consistent with a phase-transition analogy (reserve-buffered dynamics followed by an abrupt functional change when reserve falls below a critical level), but it contains no explicit percolation or network/connectome modeling, no quantitative critical-threshold analyses, and no direct early-warning signal studies.

\cite{kleiter2020} frame SPMS as a clinically identifiable 'transitional' phase with measurable thresholds used operationally (e.g., EDSS $\geq 4.0$ with pyramidal FSS $\geq 2$ and confirmed disability progression of 1.0/0.5 points), and notes natural-history conversion rates (30–50\% convert within 10–15 years; ~80\% by 20 years). It highlights limitations of EDSS thresholds (short-term confirmed increases overestimate long-term permanent disability) and suggests more objective, sensitive instruments (MSFC changes, a 0.5-point/20\% deterioration criterion, and sensor-based continuous monitoring) as potential early-warning markers of relapse-independent progression. Imaging markers (new/enlarging cortical lesions, gray-matter and cortical atrophy) and blood/CSF biomarkers of axonal damage are proposed candidates for identifying individuals at imminent risk of conversion, though blood measures are technically challenging. Cognitive measures (notably visuospatial short-term memory/learning) may discriminate RRMS vs SPMS but are confounded by early deficits in RRMS. Crucially, the authors stress that no single symptom reliably indicates progression in an individual, underscoring both the appeal of threshold-like operational definitions and their limitations. The text thus provides clinical and biomarker concepts consistent with threshold/transition thinking, while also giving counter-evidence for gradual/heterogeneous change and measurement artifacts.

\subsection*{II Network/connectome and cognitive-network percolation analyses in MS}

The task is to identify network/connectome and cognitive-network percolation analyses in MS, and evidence for threshold-like or abrupt transitions vs gradual continuum, including counter-evidence, and to extract key findings relevant to a percolation threshold idea. Most relevant evidence from wide survey follows.


\cite{lebkuecher2024} report a percolation analysis of semantic fluency networks comparing MS and neurotypical (NT) adults. Percolation integrals (where "higher percolation integrals = slower degradation = more robust network structure") were significantly larger in NT (M = 59.01, SD = 1.94) than MS (M = 55.48, SD = 1.68; t(998)=30.68, $p<.001$; large effect). Authors conclude the MS semantic networks are "less flexible and 'break apart' faster". Complementary findings: MS networks show lower overall clustering coefficient (CC), higher average shortest path length (ASPL), and higher modularity (Q), with an increased number of communities and a trend toward lower node degree; node-level clustering was significantly lower in MS (MS Mean = 0.32 vs NT Mean = 0.37). Spreading-activation simulations produced lower final activations for MS at later timepoints (timepoints 6–10, all $p < 0.01$ to $p < 0.001$), and degree and clustering predicted final activation similarly in both groups. Notably, the paper provides quantitative evidence that MS networks are less robust and fragment more rapidly under percolation, but it does not explicitly characterize a sharp percolation threshold or abrupt phase transition versus gradual degradation-no direct test or explicit identification of a critical threshold or discontinuous collapse is reported.

The thesis \cite{lall2024} describes a clique-percolation analysis of semantic (cognitive) networks in MS versus healthy controls. It used weighted clique percolation: networks were parsed into k-cliques, with clique intensity I defined as the average geometric mean of edge weights. Optimal k and I were selected by varying thresholds and identifying the point just before the giant component emerged, operationalized when the ratio of largest to second-largest community shifts from $<2$ to  $\geq 2$. The reported optimal values were k=3, I=0.13 for controls and k=3, I=0.17 for the MS network. Percolation was executed by increasing I from 0.01 to 1 stepwise, removing k-cliques below each threshold, and recording the number of connected nodes at each step. The area-under-curve (percolation integral) quantifies deterioration rate: larger integrals indicate more gradual (lower-slope) deterioration, while smaller integrals imply faster/steeper collapse. These methods therefore provide a formal way to detect threshold-like emergence/disappearance of a giant component (abrupt transition) and to quantify whether network breakdown is abrupt versus gradual via the percolation integral. The higher I threshold in MS (0.17 vs 0.13) and the integral metric are directly relevant to testing threshold-like versus continuum degeneration. The study provides methods and optimal thresholds but does not report comparative percolation integral results demonstrating whether MS shows an abrupt versus gradual transition; the slightly higher optimal I in MS (0.17 vs 0.13) suggests a shift in threshold but direct evidence of abruptness/continuum is not presented here.

In \cite{lebkuecher2024} the authors applied clique-percolation analysis to semantic fluency networks to assess vulnerability to fragmentation, running 500 iterations and computing a percolation integral per group; lower integrals indicate steeper percolation slopes and greater susceptibility to breaking apart. They compared mean percolation integrals between MS and neurotypical groups with independent-samples t-tests. Complementary network metrics showed MS networks had lower clustering coefficient (CC), higher average shortest path length (ASPL), and higher modularity (Q) - patterns consistent with reduced small-world efficiency and greater community segregation. Small-worldness values were $>2$ for both groups (MS = 2.3769, NT = 2.1473), and Network Portrait Divergence (DJS = 0.53) indicated fairly dissimilar topologies. The team controlled node counts across groups. Together, the percolation approach and the altered CC/ASPL/Q in MS provide indirect support for the idea that MS semantic networks may be more prone to abrupt fragmentation (percolation-like transitions). Counter-evidence to present hypothesis is that both groups retain small-world topology (arguing for preserved global organization), and no explicit percolation-integral results or effect sizes are reported here, limiting direct conclusions about threshold-like vs. gradual changes.

 \cite{lebkuecher2024} also describe a percolation analysis applied to cognitive (semantic fluency) networks for participants with MS versus neurotypical (NT) controls using the CliquePercolation package. Weighted semantic networks were decomposed into k-cliques (minimum k=3) and communities defined by overlapping k-cliques. Edges (k-cliques) were pruned across a threshold I varied from 0.01 to 1 to simulate progressive removal of weaker semantic links. For each threshold the number of connected nodes was recorded and the Area Under the Curve (AUC) across all I thresholds computed; from that the percolation integral was derived to quantify how quickly components separated from the giant component. A lower percolation integral indicates a steeper percolation slope (greater susceptibility to abrupt breakdown). The analysis ran 500 iterations and compared group percolation integrals via independent-samples t-tests. The paper does not report empirical results or comparisons for MS vs NT, nor does it describe any connectome (brain structural/functional) percolation analyses - only semantic/cognitive-network percolation is detailed. Thus the methods define metrics capable of distinguishing abrupt (threshold-like) vs gradual fragmentation, but no direct evidence or counter-evidence for threshold-like transitions is provided.

\cite{pagani2020} performed structural-connectivity and disconnection modeling in MS. They found that 10 network hubs were preserved across healthy controls and MS patients, and that the precuneus emerged as a consistent hub when disconnection was modeled, implicating it as a central node during GM connectivity disruption. Modeling added frontal regions and the insula among areas associated with PASAT performance, linking disconnection to cognitive changes. Crucially, the authors report that "modeling disconnection did not greatly change the picture of connectivity, suggesting that the brain is resilient to damage in MS lesions," which argues against a simple, abrupt collapse of the network (a single percolation threshold) in their empirical lesion set. At the same time, the identification of conserved hubs (especially the precuneus) suggests that targeted damage to core nodes could disproportionately impact network function-a mechanism consistent with threshold-like behavior in targeted attacks. The paper notes it used experimentally observed lesions and recommends comparing those results with simulated lesions, implying further work is needed to test for abrupt vs gradual transitions under different lesion patterns.

The study by \cite{martinezheras2023} reports widespread structural - connectivity (FA-weighted) reductions and decreased global/local graph metrics in people with multiple sclerosis (PwMS) relative to healthy controls. Overall PwMS showed reduced FA in 1686 of 1818 connections (92.7\%). Early phenotypes (CIS and RRMS) already exhibit large-scale disruptions (~80\% of connections affected). SPMS shows significant reductions across all 76 network nodes (strength, clustering, local efficiency) and lower FA in $\sim 79.9\%$ of connections; compared with PPMS, SPMS had lower FA in 303 connections (16.7\%). In contrast, PPMS showed relatively focal FA decreases (35 connections, 1.92\%). The authors used these network differences for SVM classification (FA matrices 81\% accuracy). The study does not describe any explicit percolation or threshold-modeling analyses, nor does it report statistical tests for abrupt/phase-like transitions. The empirical pattern gives mixed implications: widespread involvement from early stages suggests a continuum, while the marked worsening of nodal/global metrics in SPMS (versus PPMS/CIS-RRMS) could be interpreted as a step-like deterioration; however, direct percolation evidence or modeling of critical thresholds is not provided in this text.

 The study by \cite{tranfa2024} used conventional 3T structural MRI, automated lesion segmentation, normative tractography-based lesion masks, and the MIND morphometric similarity method to generate subject-level structural disconnection and morphometric similarity connectomes in 461 people with MS (1235 visits) and 55 controls. Network-based statistics (NBS) and NBS-predict tested associations with disease status, progression, EDSS and SDMT, and long-term confirmed disability progression (CDP). Key findings in brief: structural disconnection concentrated around thalami and cortical sensory/association hubs; EDSS related to fronto-thalamic disconnection; SDMT related to left cortico-subcortical disconnection; both network measures significantly progressed longitudinally and correlated with EDSS increases; baseline disconnection predicted long-term CDP with ~60\% accuracy; structural disconnection and morphometric similarity were positively coupled at edge/node levels. The study demonstrates network/connectome disruption, clinical correlations, longitudinal progression, and moderate predictive power. However, they do not report explicit percolation- or threshold-style analyses (no mention of abrupt transitions, critical percolation thresholds, or step-like network collapse), so they provide evidence of progressive network degradation and associations with clinical outcomes but not direct evidence for threshold-like percolation phenomena.

\subsection*{III Percolation theory limitations and empirical gaps}
\cite{kleiter2020} report that neuropathology and lesion burden evolve continuously rather than showing an abrupt phenotypic switch at transition to progressive MS: "Neuropathologically, no abrupt change in phenotype is observed" and "White and grey matter lesions as well as neuroaxonal degeneration evolve continuously." Conventional MRI lacks standardized measures predictive of SPMS conversion and "appear to be unable to predict the risk of conversion to SPMS." Structural measures that do associate with progression on group level include cortical lesion volume, gray matter atrophy (predictor of progression), spinal cord gray matter atrophy, and lesion atrophy. Fluid biomarkers show population-level signals: NfL is elevated in SPMS versus RRMS and "shows a modest positive correlation with change of EDSS over time and the rate of brain atrophy," but is not specific and "individual thresholds have not yet been prospectively determined." GFAP is elevated in SPMS and may predict long-term EDSS worsening but showed "no evident information benefit ... versus NfL" in CSF; CHI3L1 is elevated in progressive MS and combined CHI3L1+NfL elevations preceded clinical progression in one report. OCT measures (pRNFL, GCIPL) correlate with EDSS, brain atrophy and are prognostic for EDSS progression. The paper emphasizes that predictive signals are established mainly at the population level and notes persistent active lesion pathology during the transitional phase. The text does not discuss connectome percolation, critical slowing/early-warning signals, formal percolation parameter mapping, or immune avalanche statistics.
    
\cite{kleiter2020} also address markers and measurement issues around the transitional phase from RRMS to SPMS. They highlight increasing motor dysfunction as the predominant clinical sign of SPMS and note limitations of EDSS (low sensitivity to some domains, inter-rater variability, and that EDSS increases confirmed at 3–6 months overestimate permanent disability). The MSFC has objective components and a deterioration threshold (0.5 points total or 20\% in a component) that can predict later EDSS change; an operational definition of transitional MS could be MSFC progression in a relapse-free interval independent of EDSS. Continuous, rater-independent sensor and mobile monitoring (e.g., stride length at maximum speed) are proposed as sensitive early indicators of relapse-independent progression. Imaging correlates expected in transitional MS include new/enlarging cortical lesions and incipient gray matter and cortical atrophy, suggesting development of an MRI score might help detect early progression. Candidate biomarkers are markers of axonal destruction and intracerebral inflammation, but CNS molecules in blood are often extremely low (femtomolar), requiring very sensitive assays. The text also notes a insufficient number of studies on cognitive/neuropsychiatric profiles in the transitional phase and that visuospatial short-term memory/learning differ most between RRMS and SPMS.
    
\cite{filippi2020} report MRI and biofluid findings relevant to detecting progression toward secondary progressive MS (SPMS). A cortical lesion burden threshold ($>7$ lesions) at onset predicted conversion to SPMS and accelerated conversion. Global and grey-matter (GM) atrophy strongly correlate with clinical measures and predict progression, with GM atrophy rates rising substantially across disease stages (GM atrophy over 4 years: 8.1x in RRMS, 12.4x in RRMS converting to SPMS, 14x in SPMS). Deep GM atrophy is most rapid and linked to disability. Early GM microstructural worsening during the first year predicts later disability, and combining clinical plus MRI metrics can identify patients at risk years earlier. Spinal cord lesions and cervical cord atrophy are associated with greater disability. Serum neurofilament light (sNfL) is associated with EDSS and is higher in progressive phenotypes; higher baseline sNfL predicts EDSS worsening, 3-month confirmed disability, SPMS conversion, and greater future brain and spinal cord volume loss. Longitudinal sNfL rises correlate with concurrent EDSS increases (example: 10-fold sNfL increase associated with EDSS +0.53). A clear limitation noted is the need to standardize analytic methods and define pathological cut-offs. The paper does not present explicit analyses or terminology related to critical slowing, percolation/robustness curves, mapping lesion topology to percolation parameters, or avalanche/percolation statistics of immune activation.

\cite{filippi2020} cite multiple studies linking structural and fluid biomarkers to later disability/progression but does not report explicit analyses using percolation theory or early-warning (critical slowing) metrics. MRI markers of chronic active / slowly expanding lesions have been proposed and associated with disability in vivo. Lesion heterogeneity on high-field MRI correlates with MS severity. Gray matter and deep gray matter atrophy predict long-term accumulation of disability, and spinal cord (including gray matter) atrophy correlates with disability. Serum neurofilament light (NfL) is repeatedly reported as a biomarker of neuronal damage and is associated with progression, brain atrophy, and long-term outcomes. There is evidence of compartmentalized intrathecal inflammation and leptomeningeal inflammation in MS, and remyelination capacity declines with chronicity. The limitations apparent from these citations is that most findings are associative and longitudinal observational studies rather than mechanistic network- or percolation-based analyses; the paper contains no direct data on critical slowing, percolation/robustness–fragility curves, mapping lesion topology to percolation parameters (p, pc), or immune avalanche statistics. Therefore concrete claims about threshold-like dynamics or formal percolation mappings are not supported by the provided text and would require dedicated network modeling, high-temporal-resolution longitudinal biomarkers, or explicit statistical tests for early-warning signals.
    
\cite{tranfa2024} analyzed 461 people with MS across 1,235 visits and found spatially distributed subnetworks of structural disconnection and morphometric similarity disruption associated with clinical disability (EDSS) and cognition (SDMT). Global structural disconnection correlated strongly with total lesion volume (TLV; Spearman’s rho = 0.94) while morphometric similarity correlated with brain parenchymal fraction (BPF; Spearman’s rho = -0.40). Baseline structural disconnection subnetworks (but not morphometric similarity) were predictive of long-term confirmed disability progression (CDP) with a modest SVM accuracy $\sim 0.59$ (p = 0.03). Longitudinally, progressive structural disconnection (82 edges, pFWE = 0.04) and larger progressive morphometric similarity alterations (509 edges, $pFWE < 0.01$) were identified; annualized structural disconnection and morphometric similarity changes correlated weakly but significantly with annualized EDSS change ($\rho = 0.07–0.11$, $p \geq 0.02$) and remained after adjusting for TLV or BPF changes. Longitudinal models also showed EDSS worsening and whole-brain atrophy but no significant increase in global lesion burden. Limitations: the paper reports network-disconnection correlates and modest predictive performance but does not analyze percolation theory, critical slowing or early-warning signals, map lesion topology to percolation parameters (p, pc), examine immune-network avalanche statistics, nor report fluid biomarkers such as neurofilament light (NfL) showing threshold-like transitions."
    
Further, \cite{tranfa2024} report that structural disconnection and morphometric similarity alterations in MS are strongly associated with global lesion burden (TLV) and brain parenchymal fraction (BPF). Longitudinally, modest but significant correlations link annualized structural disconnection and morphometric similarity changes to EDSS worsening ($\rho \sim 0.07 - 0.11$), and progressive subnetworks of structural disconnection (mainly fronto-thalamic) and widespread morphometric-similarity change were identified. Baseline structural disconnection matrices modestly predicted long-term confirmed disability progression (CDP) with SVM accuracy $\approx 0.59$, whereas baseline morphometric similarity did not predict CDP above chance. The authors note a positive edge-level association between disconnection probability and morphometric similarity and describe this relationship as "nonlinear, multiphasic." Increase in global lesion burden over the observed timeframe was not significant. The study has a large sample and longitudinal visits, but does not report direct analyses using percolation theory, explicit critical-threshold estimates (pc) or edge-occupancy mappings, nor does it assess immune avalanche statistics, NfL, or explicit axonal-reserve-to-percolation-parameter mapping. Effect sizes for longitudinal correlations are small, and prediction accuracy for CDP is modest, indicating limits to clinical threshold detection in this dataset.

\end{document}